\documentclass[prl,twocolumn, superscriptaddress, showpacs, showkeys]{revtex4}

\usepackage{amsmath,amsfonts,amssymb,graphicx,psfrag, epsfig,rotate}
\usepackage{eucal}
\usepackage{cmmib57}

\begin{document}

\title{Thermodynamics  of the  one-dimensional {\textit{s=1}} {\textit{XXZ}} Heisenberg
                  model: \\    analytical results}

\date{\today}

\author{Onofre Rojas}
\affiliation{Departamento de F\'{\i}sica,
Universidade Federal de S\~ao Carlos,
CEP 13565-905, S\~ao Carlos - SP, Brazil}

\author{E.V. Corr\^ea Silva}
\affiliation{Instituto de F\'{\i}sica,
Universidade Federal Fluminense, Av. Gal. Milton Tavares de
Souza s/n$^{\small\textit o}$, CEP 24210-340, Niter\'oi - RJ, Brazil}

\author{Winder A. Moura-Melo}
\affiliation{Faculdades Federais Integradas de Diamantina,
 Rua da Gl\'oria n$^{\small \textit o}$ 187, CEP 39100-000,
Diamantina-MG, Brazil}

\author{S.M. de Souza}
\affiliation{Departamento de Ci\^encias Exatas, Universidade Federal de
Lavras, Caixa Postal 37, CEP 37200-000, Lavras-MG,  Brazil}

\author{M.T. Thomaz}
\email[Corresponding author: ]{mtt@if.uff.br}
\affiliation{Instituto de F\'{\i}sica,
Universidade Federal Fluminense, Av. Gal. Milton Tavares de
Souza s/n$^{\small\textit o}$, CEP 24210-340, Niter\'oi - RJ, Brazil}

\begin{abstract}

We apply the results derived by Rojas {\textit{et al.}}\cite{chain_m}  to
derive the $\beta$-expansion of the Helmholtz free energy of the
spin-1 {\textit{XXZ}} Heisenberg model up to order $\beta^5$.
The analytical expansion obtained is valid  for all phases of this
model. Our  curves of the specific heat fit well
Bl\"ote's numerical results\cite{Blote} in the high
temperature  regime.

\end{abstract}


\maketitle

The spin-$1$ $XXZ$ Heisenberg model is non-integrable and therefore
cannot be solved by the algebraic Bethe {\textit{ansatz}}
 method\cite{takahashi_L}.  Since the  70's its thermodynamics
  has been studied
  numerically\cite{Blote,neef, yamamoto93,yamamoto94,yamamoto95}.
  More recently, Yamamoto and
Miyashita\cite{yamamoto93,yamamoto94,yamamoto95} applied a Monte
Carlo method to study the specific heat, static magnetic
susceptibility and magnetization  of rings and chains of different sizes
 as a function of temperature and extended those results to the
thermodynamic limit. Bao {\textit{et al.}}\cite{termo_s1} derived
a set of self-consistent equations to approximately obtain  the
thermodynamics of this model for all range of temperature, and solved
those equations numerically.

The cummulant series\cite{domb} yields a $\beta$-expansion
of the Helmholtz free energy for exact integrable  as
well as for non-integrable models. Recently Rojas {\it et al.}\cite{chain_m}
showed for any one-dimensional chain model with periodic boundary
condition, invariance under spatial translation and interaction
between nearest neighbours that, in the
thermodynamic  limit, the coefficient of the high
temperature expansion of arbitrary order $\beta^n$ can be derived from
an auxiliary function $\varphi$.

Our aim in this work is to apply the results of Ref.~\onlinecite{chain_m}
to derive an analytical $\beta$-expansion of
the Helmholtz free energy per site of the anisotropic spin-$1$ $XXZ$
Heisenberg model  with single-ion anisotropy, up to order $\beta^5$.

  \vspace{0.5cm}

The hamiltonian of the spin-$1$ $XXZ$ Heisenberg model with
anisotropy\cite{wang} is

\begin{equation}
H = \sum_{i=1}^N J \left( S^{+}_{i} S^{-}_{i+1} +
	S^{-}_{i} S^{+}_{i+1} \right)
	+ \Delta S^{z}_{i} S^{z}_{i+1}
	- h S^{z}_{i}
	+ D (S^{z}_{i})^2~,    \label{1a}
\end{equation}

\noindent where $N$  is the number of sites in the periodic chain,
$\Delta$ the anisotropy constant, $h$ the external magnetic
field in the $z$-axis and $D$ the single-ion anisotropy parameter.
We have
$S_i^{\pm} \equiv \frac{1}{\sqrt{2}} \left( S_i^{x} \pm  i \ S_i^{y} \right)$.
We do not fix the sign of any constant in eq.(\ref{1a}).

Applying the results of Ref.~\onlinecite{chain_m} to the
spin-$1$ $XXZ$ Heisenberg model with single-ion anisotropy,
we obtain  its Helmholtz free energy  ${\mathcal W}(\beta)$
up to order $\beta^5$:

\begin{eqnarray}  
{\mathcal W}(\beta) &=&
	-{\frac {\ln (3)}{\beta}}
	+ \frac{2}{3}\,D
	+
\nonumber \\
& + &
	\left (
	-\frac{1}{9}\,{D}^{2}
	-\frac{4}{9}\,{J}^{2}
	-\frac{1}{3}\,{h}^{2}
	-\frac{2}{9}\,{\Delta}^{2}
	\right )\beta
	+
\nonumber \\
& + &
	\left(
	-{\frac {1}{81}}\,{D}^{3}
	+\frac{4}{9}\,\Delta\,{h}^{2}
	+{\frac {4}{27}}\,{\Delta}^{2}D
	-{\frac {4}{27}}\,{J}^{2}D
	+ \right.
\nonumber \\
&  &
	\left.
	+\frac{1}{9}\,{h}^{2}D
	-\frac{1}{9}\,{J}^{2}\Delta
	\right){\beta}^{2}
	+
\nonumber \\
& + &
	\left (
	\frac{1}{36}\,{h}^{4}
	-{\frac {1}{54}}\,{\Delta}^{4}
	+{\frac {1}{324}}\,{D}^{4}
	+{\frac {13}{81}}\,{J}^{2}{\Delta}^{2}
	+ \right.
\nonumber \\
&  &
	+ {\frac {1}{54}}\,{h}^{2}{D}^{2}
	+{\frac {7}{162}}\,{J}^{4}
	- {\frac {8}{27}}\,\Delta\,{h}^{2}D
	+{\frac {5}{27}}\,{J}^{2}{h}^{2}
	+
\nonumber \\
&  &
	\left.
	-{\frac {10}{27}}\,{\Delta}^{2}{h}^{2}
	+ \frac{1}{27}\,{J}^{2}{D}^{2}
	\right) {\beta}^{3}
	+
\nonumber \\
& + &
	\left (
	{\frac {7}{243}}\,{J}^{2}{D}^{3}
	+\frac{1}{18}\,{J}^{4}\Delta
	+{\frac {1}{972}}\,{D}^{5}
	+ \right.
\nonumber \\
&  &
	-\frac{1}{27}\,D{J}^{2}{h}^{2}
	+{\frac {1}{81}}\,{J}^{2}\Delta\,{D}^{2}
	-{\frac {1}{81}}\,D{\Delta}^{4}
	+
\nonumber \\
&  &
	-{\frac {11}{162}}\,D{J}^{2}{\Delta}^{2}
	-{\frac {4}{243}}\,{\Delta}^{2}{D}^{3}
	+{\frac {22}{81}}\,{h}^{2}{\Delta}^{3}
	+
\nonumber \\
& &
	-{\frac {10}{27}}\,{J}^{2}\Delta\,{h}^{2}
	-\frac{1}{36}\,D{h}^{4}
	+{\frac {8}{27}}\,{\Delta}^{2}{h}^{2}D
	+
\nonumber \\
& &
	-{\frac {1}{162}}\,{D}^{3}{h}^{2}
	+ \frac{1}{36}\,{J}^{2}{\Delta}^{3}
	+{\frac {13}{162}}\,D{J}^{4}
	+
\nonumber \\
& &
	\left.
	-{\frac {4}{27}}\,{h}^{4}\Delta
	\right ){\beta}^{4}
	+
	\nonumber
\end{eqnarray}
\begin{eqnarray} \label{3}
& + &
	\left (
	-\frac{1}{18}\,{D}^{2}{h}^{2}{J}^{2}
	+{\frac {16}{81}}\,D{h}^{4}\Delta
	-{\frac {2}{81}}\,{D}^{2}{h}^{2}{\Delta}^{2}
	+ \right.
\nonumber \\
& &
	+{\frac {5}{243}}\,{D}^{2}{\Delta}^{4}
	-{\frac {77}{486}}\,{h}^{2}{\Delta}^{4}
	+ {\frac {13}{1620}}\,{D}^{2}{J}^{4}
	+
\nonumber \\
&  &
	-{\frac {31}{324}}\,{J}^{4}{h}^{2}
	-{\frac {5}{1944}}\,{D}^{4}{h}^{2}
	-{\frac {7}{14580}}\,{D}^{4}{J}^{2}
	+
\nonumber \\
& &
	-{\frac {53}{4860}}\,{J}^{2}{\Delta}^{4}
	+{\frac {29}{81}}\,{h}^{4}{\Delta}^{2}
	-{\frac {1241}{14580}}\,{\Delta}^{2}{J}^{4}
	+
\nonumber \\
& &
	-{\frac {19}{324}}\,{J}^{2}{h}^{4}
	+{\frac {1}{648}}\,{D}^{2}{h}^{4}
	-{\frac {1}{729}}\,{D}^{4}{\Delta}^{2}
	+
\nonumber \\
& &
	+{\frac {8}{243}}\,{h}^{2}{D}^{3}\Delta
	-{\frac {52}{243}}\,{h}^{2}D{\Delta}^{3}
	+{\frac {26}{81}}\,{J}^{2}{h}^{2}{\Delta}^{2}
	+
\nonumber \\
& &
	-{\frac {76}{1215}}\,{J}^{2}{D}^{2}{\Delta}^{2}
	+{\frac {2}{1215}}\,\Delta\,{D}^{3}{J}^{2}
	-{\frac {2}{81}}\,D{J}^{2}{\Delta}^{3}
	+
\nonumber \\
& &
	+{\frac {2}{81}}\,D\Delta\,{J}^{4}
	-{\frac {7}{87480}}\,{D}^{6}
	-{\frac {13}{3240}}\,{h}^{6}
	+
\nonumber \\
& &
	\left.
	+ {\frac {173}{43740}}\,{\Delta}^{6}
	-{\frac {131}{43740}}\,{J}^{6}
	+{\frac {14}{81}}\,{h}^{2}D\Delta\,{J}^{2}
	\right ){\beta}^{5}
	+
\nonumber \\
& + &
	{\mathcal O}(\beta^6).
\label{HelmholtzEn}
\end{eqnarray}

\noindent We point out that this analytic expansion is valid for
{\textit{any}} arbitrary set of values of the parameters $J$, $\Delta$,
$D$ and $h$ in the high temperature regime. Therefore it
applies equally well to any phase of the one-dimensional spin-1 $XXZ$
Heisenberg model\cite{chui,glaus,solyom}.
The coefficients for each order in $\beta$ in the expansion (\ref{HelmholtzEn})
are exact.

In addition, from eq.(\ref{3}) we see that the thermodynamic properties of this
model are insensitive to the sign of the constants $J$ and $h$.
 From now on we take $J\geq 0$. By redefining  all
the parameters of the hamiltonian (\ref{1a}),
eq.(\ref{3}) becomes an expansion in powers of $(J\beta)^n$.

A beautiful numerical work by Bl\"ote\cite{Blote}
tabulates numerical values for the specific heat at varied
temperatures (including the high temperature region)
  for the $s=1/2$ and $s=1$ Heisenberg models. In all his calculations Bl\"ote
 has set $h=0$. To verify the correctness of our expansion (\ref{HelmholtzEn}) we
compare the analytical expression for the specific heat derived
from it with the numerical results of Ref.~\onlinecite{Blote}.
Comparison of the hamiltonian (1) in
 Ref.~\onlinecite{Blote} and ours, given by eq.(\ref{1a}),
 yields the following
 correspondence of parameters: $J_{\footnotesize \parallel}= - \Delta/2$ and
 $J_{\footnotesize \bot}= - J/2$. In Fig.~\ref{Figura1} we compare
  our analytical expression  for the specific heat
 derived from eq.(\ref{3}) for anisotropic Heisenberg models in the
 absence of single-ion anisotropy ($D=0$)
 with different values of $J$ in the antiferromagnetic phase.
 In the high temperature region our curves fit well the
  numerical data.

It is certainly important to verify that eq.(\ref{HelmholtzEn}) applies equally
well to the ferromagnetic phase of the spin-1 Heisenberg model. To show this
we have Fig.~\ref{Figura2} where we plot the specific heat, in the high 
temperature region, of the isotropic  Heisenberg model with different
values for the single-ion parameter $D$ in this phase. As in Fig.~\ref{Figura1}, our
curves  in Fig.~\ref{Figura2} fit pretty well Bl\"ote's
numerical results in the high temperature region.

Our analytical results for the specific heat,
obtainable from  (\ref{HelmholtzEn}),
fully agrees in its isotropic limit with that of
section 5.2 of Ref.~\onlinecite{Blote}
in the high temperature limit. We also recover
the high temperature static magnetic susceptibility
for the isotropic Heisenberg model [eq.(4.5a) of Ref.~\onlinecite{yamamoto94}].
However, we do not agree with the high temperature limit of
the correlation function for the z-component of spin between nearest
neighbors $\langle S_0^z S_1^z \rangle$
of Ref.~\onlinecite{termo_s1}. The correct
limit is $\langle S_0^z S_1^z \rangle \approx - \frac{4}{9}\Delta \beta$.

\vspace{0.5cm}

In summary, in this letter we applied the
 method of Ref.~\onlinecite{chain_m}
to a non-integrable spin-1 $XXZ$ Heisenberg model with
single-ion anisotropy in the presence of an external
magnetic field in the $z$-direction. We obtained
the analytical $\beta$-expansion of the Helmholtz free energy  per site
of the model up to order $\beta^5$. Each coefficient in
expansion (\ref{3}) is exact and valid for {\textit{any}} phase
of the model.

 Our curves in the high temperature region fit well
 Bl\"ote's  numerical results for the spin-1 $XXZ$ Heisenberg model.
We recover the few analytical results
known in the literature\cite{Blote, yamamoto94} about
the isotropic spin-1 Heisenberg model. We correct the high
temperature limit of  $\langle S_0^z S_1^z \rangle$ of
Ref.~\onlinecite{termo_s1}.  For the first time in the literature
we have analytical results easily handled for the anisotropic
spin-1 Heisenberg model with single-ion anisotropy
 and in the presence of an external magnetic field.

Finally we should mention that our calculation procedures
have been implemented in the symbolic computational language
{\texttt Maple}\cite{Eduardo}.
Currently, refinements are being made so that the expansion of
the Helmholtz free energy (\ref{3}) for the $s=1$ Heisenberg model can be
extended to higher orders in $\beta$.

An enlarged version of this letter  will be published elsewhere.

\begin{acknowledgments}
 O. R. thanks CLAF for financial support. E.V.C.S.
 thanks CNPq for financial support. S.M. de S. thanks
  FAPEMIG for partial financial support.
 M.T.T. thanks CNPq and FAPERJ for partial financial
  support. WAMM acknowledges FAPEMIG and CNPq for partial
  financial support.
\end{acknowledgments}

\newcommand{\ppsfrag}[2]{\psfrag{#1}{\small {#2}}}

{

\ppsfrag{TituloFigura4}{\hspace*{-1cm}Specific Heat }
\ppsfrag{Eixox}{\hspace*{1cm}$\beta$}
\ppsfrag{Eixoy}{$C_v$}
\ppsfrag{Aux1}{\hspace*{5mm}$J=0$}
\ppsfrag{Aux2}{\hspace*{-7mm}$J=1$}
\ppsfrag{Aux3}{\hspace*{-5mm}$J=1.8$}
\ppsfrag{Aux4}{{\begin{tabular}{l} $\Delta=+2$ \\ $D=0$ \\ $h=0$ \end{tabular}}}

\begin{figure}[htb]
\includegraphics[width=8.6cm,height=8.6cm]{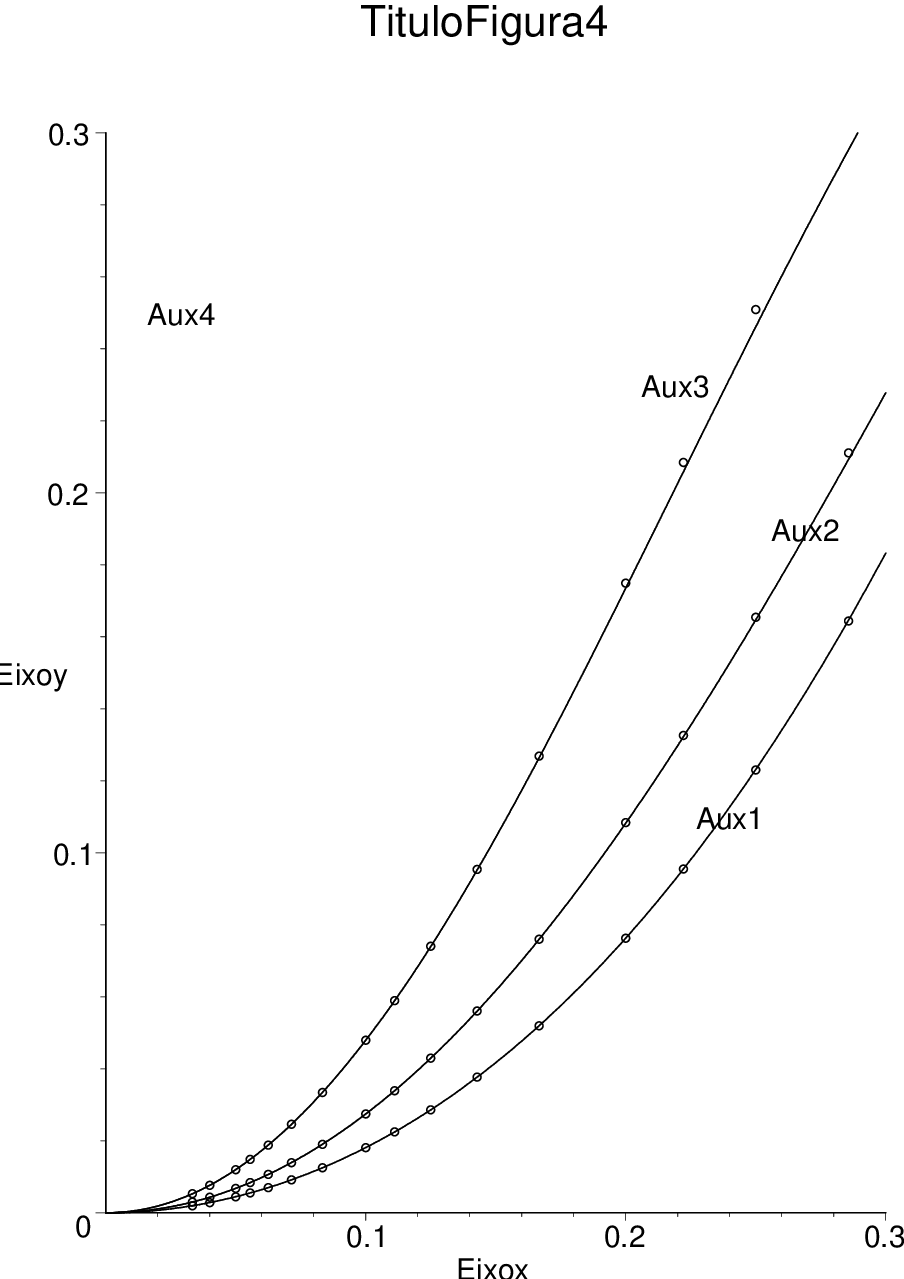}
\caption{\label{Figura1} Comparison of our analytic expression for the
specific heat and Bl\"ote's numerical result\cite{Blote}
for the anisotropic Heisenber model with different anisotropies
 and $h=0$.  Solid lines stand for the
specific heat of our  analytical result and dotted
lines correspond to data from
Ref.~\onlinecite{Blote}.}


\ppsfrag{TituloFigura5a}{\hspace*{-1cm}  Specific Heat }
\ppsfrag{Eixox}{\hspace*{1cm}$\beta$}
\ppsfrag{Eixoy}{$C_v$}
\ppsfrag{Aux1}{$D=0.5$}
\ppsfrag{Aux2}{\hspace*{-2mm}$D=2$}
\ppsfrag{Aux3}{$D=5$}
\ppsfrag{Aux4}{{\begin{tabular}{l} $J=-2$ \\ $\Delta=-2$ \\ $h=0$ \end{tabular}}}

\vspace{1cm}
\includegraphics[width=8.6cm,height=8.6cm]{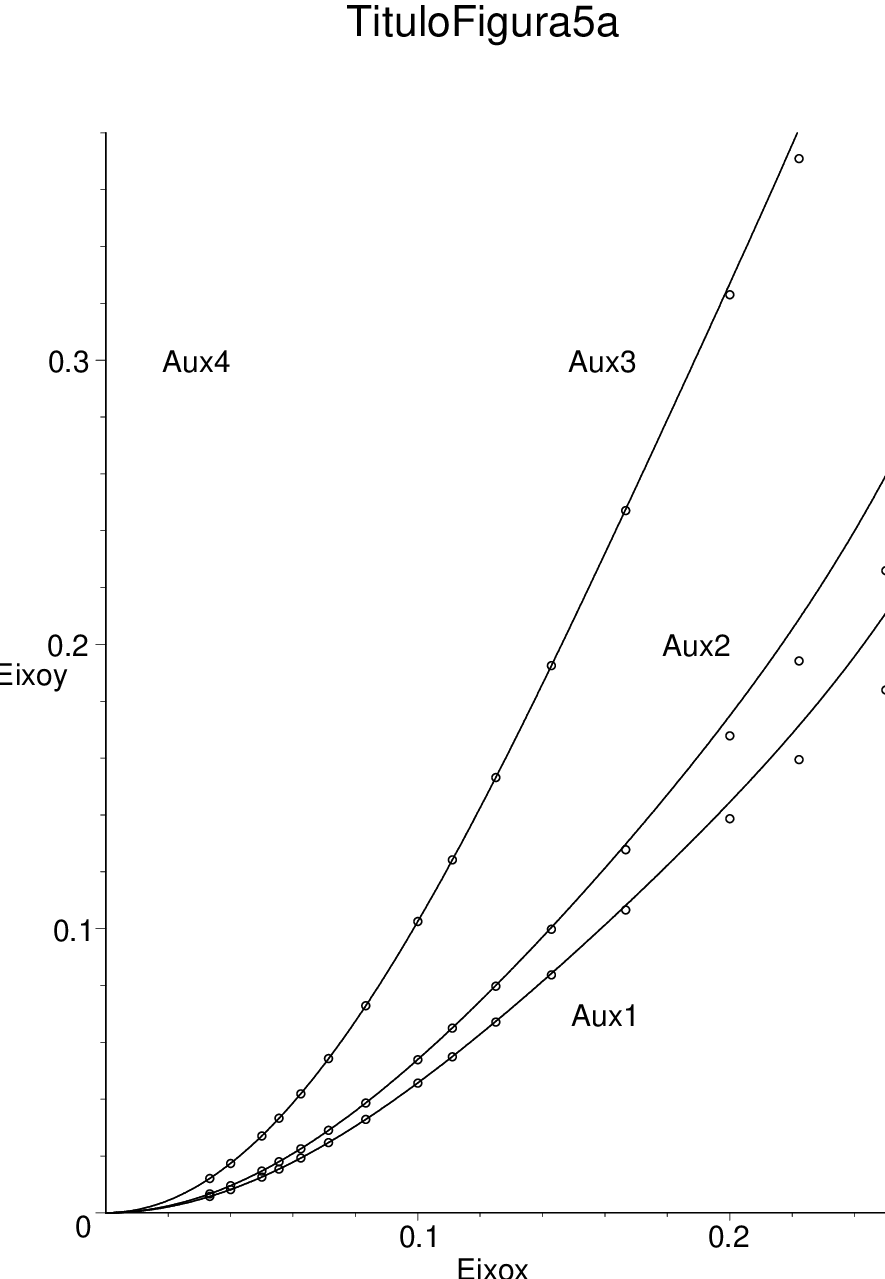}
\caption{\label{Figura2}Comparison of our analytic expression for the
specific heat and Bl\"ote's numerical results\cite{Blote}
for the isotropic Heisenberg model with different  single-ion
parameters and  $h=0$.
Solid lines stand for our results and dotted lines
correspond to data from \cite{Blote}.}
%
%
\end{figure}
}

\end{document}